\documentstyle[12pt]{article}

\begin{document} 
\begin{flushright} {OITS 660\\
September 1998}
\end{flushright}
\vspace*{1cm}

\begin{center} 
{\large {\bf Entropy Index as a Measure of Heartbeat
Irregularity}}
\vskip .75cm
 {\bf  RUDOLPH C. HWA }
\vskip.5cm
 {Institute of Theoretical Science and Department of Physics\\ University of
Oregon, Eugene, OR 97403-5203, USA\\E-mail: hwa@oregon.uoregon.edu}
\end{center}

\begin{abstract}
A method is proposed to analyze the heartbeat waveform that can yield
a reliable characterization of the structure after only a few pulses.  The
measure suggested is entropy index that is related to the one found
effective in describing chaotic behaviors in a wide variety of physical
systems.  When applied to the ECG data that include ventricular
fibrillation, the index is shown to change drastically within a few
pulses.  Wavelet analysis is used to exhibit different scaling behaviors
in different phases.
\end{abstract}

Concepts from the theory of nonlinear dynamics and statistical physics
have been applied to the study of nonstationary time series, such as
human heartbeats \cite{cp,cp2,gv,st} and brain electrical
activities\cite{hk,kl}.  Those analyses generally involve the use of data
recorded over a long period of time.  In this paper we propose a method
of analysis that needs only a short duration of the time series
data, as short as, say, ten heartbeats, for example.  Conceptually, the
method is interesting because the proposed measure is related to the
study of fluctuations in a diverse range of physical problems, e.g., hadron
production in high-energy collisions \cite{zc,sw}, classical chaotic
systems \cite{zc2}, and phase transition \cite{rh}.  The underlying
universality in all those applications is rooted in the attempt to quantify
the fluctuations of spatial patterns, for which an effective measure found
is the entropy index  \cite{zc,zc2}.  The application to the ECG time series,
as we shall discuss here, provides a diagnostic tool that is both simple
and efficient.  

A major effort initiated by physicists to analyze the human heartbeat
time series is to study the fluctuation of time intervals between the
$R$ pulses.  The study of such fluctuations is motivated by the possible
analogy with critical behaviors in statistical systems, where fluctuations
at all length (time) scales occur.  While that is certainly an interesting
area of investigation, the data required run in excess of $10^4$
heartbeats \cite{gv,st}.  There are, however, a great deal of information
about the heartbeat time series that is discarded when the focus is only
on the interbeat time intervals.  As is well known, the structure of the
time series between beats changes drastically when a heart goes into
fibrillation \cite{dm}.  The question is how to quantify that structure in
an efficient manner so that the numerical value of an appropriate
measure can be determined after a few beats.  Of course, there is no
need for such a measure if one has at hand the data for both before and
during fibrillation, anymore than the need for a smoke detector when a
house is actually on fire.  However, the availability of a numerical
measure of the cardiac activity is clearly a useful tool, especially for
patients with irregular behaviors of the heart.

The analysis that we propose has its origin in the study of spatial
patterns associated with the final state of particles in momentum space
detected at the end of each event in high-energy collisions \cite{zc}.  For
each event the factorial moments are used to describe the pattern; those
moments have the virtue of filtering out the statistical fluctuations
\cite{ab}.  The nature of the fluctuations of those moments from
event to event is quantified by an index $\mu$, which is larger
when the fluctuation is larger.    For a heartbeat time series, we
partition it into many segments of  short duration (e.g., 2 sec),  regard
each segment as a pattern, and characterize each pattern by studying the
fluctuations from bin to bin.  Since statistical fluctuation does not have
the same meaning in the heartbeat problem as for particle production,
we shall not use the factorial moments.  In their place we shall employ
the wavelet analysis
\cite{id,gk,mg}, which is natural for a problem that has sharp spikes and
low bumps.  The corresponding entropy index is then a measure of the
fluctuation of the normalized wavelet coefficients at various scales of
resolution.

In Fig.\ 1 we show the digitized electrocardiogram data that we shall
analyze.  The data were provided by Minh \cite{md},
recorded at the Stanford University Medical School, when a patient's
heart went into ventricular fibrillation, followed by a defibrillation
process.  The three phases (normal, abnormal and recovery) are clearly
identifiable visually in Fig.\ 1.  In Fig.\ 2 are shown in more detail the
structures between the pulses in the normal and abnormal phases. 
Evidently, the minor peaks and dips between the major spikes (called $R$
``waves'' \cite{dm}) behave very differently in the two phases.  To
capture and characterize those differences is therefore our task.

In the digitized data of Fig.\ 1 there are roughly 240 points between two
successive $R$ pulses in the normal phase, which spans about 3600
points.  There are approximately 2500 points in the abnormal phase. 
Given the data, we divide the time series into segments of 512 points
each, calling each segment $S_n$, with $n = 1, \cdots, 7$ belonging to
the normal phase, $n = 8, \cdots 12$ to the abnormal phase, and the rest
$n = 13, \cdots, 16$ to the recovery phase.  The number of segments in a
particular phase is not important, since the fluctuation of the patterns
from segment to segment within a phase is not large.  Thus for an
ordinary  time series that does not include a change of phase, one may
have only 10 - 20 heartbeats in a diagnostic test.  That should be
sufficient for the proposed analysis to be performed.  

Note that each segment has $2^9$ points.  The importance of that number
to be an integer power of $2$ will become self-evident, as we perform
the wavelet analysis whose resolution improves by powers of $2$.  To
have more points in a longer segment will not improve the analysis
because the information to be extracted lies with the shape of the
waveform between and around the pulses.  To have less points would
shorten the range of resolutions and inhibit the establishment of a
convincing scaling behavior.

Let the Haar wavelet $\psi^H_{jk}(t)$ be defined by
\begin{eqnarray}
\psi^H_{jk}(t) = \psi^H\left(2^j t - k\right) \quad ,
\label{1}
\end{eqnarray}
where $\psi^H(t)= 1$ for $0 \leq t < 1/2$, = $-1$ for $1/2 \leq t < 1$, and
$= 0$ otherwise.  For any scalar function $f(t)$ defined in $0 \leq t \leq
1$, the wavelet coefficient after a discrete wavelet transform is
\begin{eqnarray}
w_{jk} = \left(\psi^H_{jk}, f\right) = \int dt \, \psi^H_{jk}(t) f (t)
\quad .
\label{2}
\end{eqnarray}
By virtue of the properties of $\psi^H_{jk}(t)$, which is zero for $t$
outside the interval $\left[k2^{-j}, (k+1) 2^{-j}\right],\\  w_{jk}$ selects a
narrow sector of $f(t)$ that depends on the scale factor $j$ and shift
variable $k$.  Thus with appropriate values of $j$ and $k$, $w_{jk}$ can
identify spikes in $f (t)$.

We use (\ref{2}) to analyze the waveforms of various segments $S_n$ of
our ECG time series separately.  Since each segment has 512 points, we
consider the range of $j$ values from 0 to 8, so that at the highest
resolution two neighboring points are resolved by the transform.  The
shift $k$ can vary from 0 to $2^j - 1$.  For the purpose of our use of the
wavelet coefficients below, we want to avoid negative values by taking
the absolute value of the transform, i.e., for the $n$th segment,
\begin{eqnarray}
w^{(n)}_{jk} = \left|\left(\psi^H_{jk}, S_n\right)\right| 
\quad ,
\label{3}
\end{eqnarray}
where we have mapped the 512 points on the time axis to the interval
$0 \leq t \leq 1$.  With the definition in (\ref{3}), the average (over all
$k$ at fixed $j$)
\begin{eqnarray}
\left<w^{(n)}_{jk}\right> = 2^{-j} \sum^{2^j-1}_{k = 0}w^{(n)}_{jk} 
\label{4}
\end{eqnarray}
is always positive definite.  We now can define a normalized wavelet
coefficient
\begin{eqnarray}
z^{(n)}_{jk} = w^{(n)}_{jk}/\left<w^{(n)}_{jk} \right> \quad ,
\label{5}
\end{eqnarray}
which measures the fluctuation of $w^{(n)}_{jk}$ from the average.  This
is an important step that combines both the local and global properties
of the waveform in a segment, since $z^{(n)}_{jk} $ is sensitive to the
values of  $w^{(n)}_{jk}$ in all bins. Moreover, note that in the ratio
(\ref{5}) the normalization of the Haar wavelet
$\psi^H_{jk}(t)$ defined in (\ref{1}) is unimportant.

In order to quantify the fluctuations of $z^{(n)}_{jk}$ from bin to bin,
we now define 
\begin{eqnarray}
K^{(n)}_j = \left<z^{(n)}_{jk}\, {\rm ln} \, z^{(n)}_{jk} \right> \quad ,
\label{6}
\end{eqnarray}
where the angular brackets denote an average over $k$ as defined in
(\ref{4}).   $K^{(n)}_j$ is not far from being the
entropy.  If we define $p^{(n)}_{jk} =
2^{-j}z^{(n)}_{jk}$ with $\sum_k p^{(n)}_{jk} = 1$, we can define the
entropy as
\begin{eqnarray}
S^{(n)}_j = - \sum_k p^{(n)}_{jk} {\rm ln} \, p^{(n)}_{jk} \quad .
\label{7}
\end{eqnarray}
It  then follows that
\begin{eqnarray}
S^{(n)}_j = j {\rm ln} \,  2 - K^{(n)}_j \quad.
\label{8}
\end{eqnarray}

In the study of problems of this type that have fluctuations at all scales,
we look for scaling behavior as an organizing feature.  The quantities
that are expected to possess scaling behaviors are the moments
\begin{eqnarray}
C^{(n)}_p(M) = \left<(z^{(n)}_{jk})^p \right> \quad ,
\label{9}
\end{eqnarray}
where the dependence on $j$ may be expressed in terms of the number
of bins, $M$, via $M = 2^j$.  Clearly, we have from (\ref{6})
\begin{eqnarray}
K^{(n)}_j =  \left.{d \over dp} C^{(n)}_p\right|_{p=1} \quad .
\label{10}
\end{eqnarray}
Thus, if $C^{(n)}_p$ has the scaling behavior
\begin{eqnarray}
 C^{(n)}_p(M) \propto M^{\psi^{(n)}_p} \quad ,
\label{11}
\end{eqnarray}
as the resolution is increased (i.e., higher $M$), then it follows from
(\ref{10}) and (\ref{11}) that
\begin{eqnarray}
K^{(n)}_j \propto \mu^{(n)} {\rm ln} \,  M =  \mu^{(n)} j {\rm ln} \, 
2\quad ,
\label{12}
\end{eqnarray}
where $\mu^{(n)} =  \left.{d \over dp}\psi^{(n)}_p\right|_{p=1}$.  Our
entropy index is defined by
\begin{eqnarray}
\sigma^{(n)} = 1-\mu^{(n)} \quad,
\label{13}
\end{eqnarray}
 which follows naturally from (\ref{8}) and (\ref{12}).

The procedure for analyzing the data should now be clear and
straightforward.  For each segment $S_n$, use (\ref{3}) - (\ref{6}) to
determine $K^{(n)}_j$ for $j = 0, \cdots, 8$.  As an illustration of the
result, we show in Fig.\ 3 $K^{(n)}_8$ vs $n$ for $j = 8$.  Evidently,
$K^{(n)}_8$ is quite stationary at around 2.1 for $n = 1, \cdots, 7$, which
are the segments in the normal phase. Then at $n = 8$, $K^{(n)}_8$ drops
down to below 1 and stays below for the remaining segments of the
abnormal phase, $n = 8, \cdots, 12$.  The fluctuations
from segment to segment are not significant within one or the other of
the two phases.  Since $j = 8$ is the highest resolution that the data
allow, it provides the most dramatic changes of $K^{(n)}_j$ in the
transitions between phases.  At lower $j$ the spikes in the waveform
are smeared by the wavelet transform, with the consequence that the
changes in
$K^{(n)}_j$ between phases become less pronounced.  In that sense
$K^{(n)}_8$ itself can serve as a measure of cardiac regularity.

To capture the information contained in $K^{(n)}_j$ at lower $j$, we
investigate the scaling behavior (\ref{11}), which implies a linear
dependence of $K^{(n)}_j$ on $j$, as given in (\ref{12}).  To have an
average measure over all segments in a particular phase, we define the
following averages
\begin{eqnarray}
K^N_j =  {1  \over  7}\,  \sum^7_{n = 1} K^{(n)}_j \,  , \qquad K^A_j =  {1 
\over  5}\,  \sum^{12}_{n = 8} K^{(n)}_j
\label{14}
\end{eqnarray}
for the normal and abnormal phases, respectively.  Beginning with
$n = 13$, the recovery phase commences.  Since $K^{(n)}_j$ in the
recovery phase is time dependent, a similar average is less meaningful,
although it can analogously be defined.  In Fig.\ 4 we show $K^N_j$ and
$K^A_j$ vs
$j$, which clearly exhibit linear behavior.  The straightline fits give 
\begin{eqnarray}
\mu _{N,A} =  {1  \over   {\rm ln} \,  2}\,  {\partial \over \partial j}
K^{N,A}_j
\quad .
\label{15}
\end{eqnarray}
Their numerical values obtained are $
\mu _N = 0.54$ and $\mu _A = 0.12$. The corresponding values of the
entropy indices are then
\begin{eqnarray}
\sigma_N=0.46, \qquad\qquad \sigma_A=0.88.
\label{16}
\end{eqnarray}
 If the same procedure is followed for the
recovery phase, the corresponding entropy index is $\sigma _R = 0.58$,
which is a coarse summary of the transitory change from $\sigma _A$
back to $\sigma _N$.  Eq. (\ref{16}) exhibits the numerical result of this
work.  Whereas $\sigma_A$ may vary, depending on the nature
of the cardiac abnormality, $\sigma_N=0.46$ can be regarded as the
standard number for a normal heartbeat.  To register the state of
cardiac health in terms of $\sigma$ is clearly useful.

The increase of $\sigma$ in the transition from the normal to abnormal
phase signifies the increase of disorder in the waveform.  That is a
feature that  is visually obvious from Figs. 1 and 2. We now have a
quantitative measure of that disorder.  In the normal phase   the
fluctuation from bin to bin is relatively small despite the large, but
regular, spikes, whereas the irregularity in the abnormal phase
generates large fluctuations.

It is pertinent to remark that from the fluctuations of beat-to-beat
intervals studied over very long periods it has been inferred that the
normal heartbeat is chaotic \cite{csp}.  Since stochastic disorder is not
the same as chaotic behavior, there is no obvious conflict between that
conclusion and ours.  Nevertheless, it would be useful to point out here
the possible source of the difference in interpretations.  Because the
emphasis in this paper is on the characterization of ECG waveforms in
short periods, we have considered only a few segments with detailed
analysis of the bin-to-bin fluctuations.  To study chaotic behavior, we
would have to consider segment-to-segment fluctuations over a long
period.  That happens to be the type of analysis done earlier with the
$\mu$ index (for event-to-event fluctuations) for both classical-chaotic
time series \cite{zc2} and quantum systems involving particle production
\cite{zc}.  In fact, it was found that $\mu$ can play the role of the
Lyapunov exponent.  Applying similar method to data collected
over very long periods, it should be possible to make more elaborate
analysis, not just on the beat-to-beat intervals, but on the fluctuation of
the interpulse structure.  

To illuminate the dual properties of stochasticity and chaoticity would be
highly interesting.  Here we present only the method of analysis that
can quantify the disorder aspect of the ECG  waveform structure. 
Multichannel analysis and finding predictive signatures in correlated
data are examples of other problems well worth further investigation.

I am grateful to Dr. Minh Duong-Van for getting me interested in cardiac
problems and for letting me use the heartbeat data that he collected. 
This work was supported, in part, by U.\ S.\ Department of Energy under Grant
No. DE-FG03-96ER40972.

\section*{Figure Captions}

\begin{description}
\item[Fig.\ 1]Time series of human heartbeats that includes a period of
ventricular fibrillation. Each unit on the horizontal scale is 1/256 s;
the vertical scale has arbirary unit.
\item[Fig.\ 2]Details of Fig.\ 1 in (a) the normal phase, and (b) the
abnormal phase.
\item[Fig.\ 3]$K^{(n)}_j$ at $j = 8$ for various segments $S_n$.
\item[Fig.\ 4]Scaling behaviors of $K_j$ for the normal and abnormal
phases.
\end{description}

\end{document}